# The Evolution of Natural Cities from the Perspective of Location-Based Social Media

Bin Jiang and Yufan Miao

Department of Technology and Built Environment, Division of Geomatics
University of Gävle, SE-801 76 Gävle, Sweden
Email: bin.jiang@hig.se, yufanmiao@gmail.com

**Abstract**
This paper examines the former location-based social medium Brightkite, over its three-year life span, based on the concept of natural cities. The term 'natural cities' refers to spatially clustered geographic events, such as the agglomerated patches aggregated from individual social media users' locations. We applied the head/tail division rule to derive natural cities, based on the fact that there are far more low-density locations than high-density locations on the earth surface. More specifically, we generated a triangulated irregular network, made up of individual unique user locations, and then categorized small triangles (smaller than an average) as natural cities for the United States (mainland) on a monthly basis. The concept of natural cities provides a powerful means to develop new insights into the evolution of real cities, because there are virtually no data available to track the history of cities across their entire life spans and at very fine spatial and temporal scales. Therefore, natural cities can act as a good proxy of real cities, in the sense of understanding underlying interactions, at a global level, rather than of predicting cities, at an individual level. Apart from the data produced and the contributed methods, we established new insights into the structure and dynamics of natural cities, e.g., the idea that natural cities evolve in nonlinear manners at both spatial and temporal dimensions.

## 1. Introduction

Once upon a time, there were no cities, only scattered villages. Over time, cities gradually emerged through the interaction of people or residents; similarly, large or mega cities evolve through the interaction of cities or people. This is a thought experiment mentioned in Jiang (2014), in which he argued that many geographic phenomena such as urban growth are essentially unpredictable. Many models in the literature that claim to be able to predict urban growth are in effect for short-term prediction like the weather forecast; weather forecast beyond five days is essentially unforecastable (Bak 1996). A typical city may have hundreds of years of history, making it nearly impossible to track its growth quantitatively because of a lack of related data. More important, a city grows within a system of cities; one cannot understand a city's growth without considering other related cities. In this paper, we illustrate that emerging social media provide an unprecedented data source for studying the evolution of natural cities (c.f., Section 2 for the definition), and subsequently for better understanding structure and dynamics of real cities. Location-based social media, sometimes termed as location-based social networks, such as Flickr, Twitter, and Foursquare (Traynor and Curran 2012, Zheng and Zhou 2011) refer to a set of Internet-based applications founded on Web 2.0 technologies and ideologies that allow users to create and exchange user-generated content. Location-based social media can act as a proxy of real cities (or human settlements in general) and provide better understanding of their underlying structure and dynamics.

Not a long ago, there were no social media, only scattered home pages and bulletin board systems created and maintained by individuals and institutions (Boyd and Ellison 2008, Kaplan and Haenlein 2010). In the era of Web 1.0, geographic locations were not an issue. However, with Web 2.0, geographic locations have been becoming an important feature of social media. Almost all social media allow users to tag their geographic locations, often at the level of meters, when sharing and exchanging user-generated content. Location-based social media enable users to track individual historical trajectories, their friends, and even the growth of social media. Unlike with conventional cities, the trajectories of social media are well documented by the hosting companies; and unlike conventional census data, social media data is defined at individual level, often at very fine spatial and temporal scales. Data can be obtained using crawling techniques or through the social media's officially released application programming interfaces. This study aimed to showcase how social media's time-stamped location data can be utilized to study the evolution of natural cities, and thus, providing new insights into the underlying structure and dynamics of real cities.

The contribution of this paper can be seen from the three aspects: data, methods, and new insights. This study produced a large amount of data regarding natural cities from the former social medium Brightkite during its entire 31-month life span. The resulting data has significant value for further study of city growth and allometric relationship between populations and physical extents (data available here: https://sites.google.com/site/naturalcitiesdata/). We drew upon a set of fractal or scaling oriented methods to characterize natural cities. These unique methods help create new insights into the evolution of natural cities as well as that of real cities. For example, natural cities demonstrate a striking nonlinear property, spatially and temporally (see Section 4). Moreover, the evolution of natural cities can provide better understanding of social media from a unique geospatial perspective.

This study provides new perspectives, as well as different ways of thinking, to the study of cities and city growth in the era of big data (Mayer-Schonberger and Cukier 2013). We did not adopt conventional census data, but rather the emerging georeferenced social media data; we did not adopt conventional geographic units or boundaries that are imposed from the top down by authorities, but rather the naturally defined concept of natural cities, to avoid statistical bias out of the modifiable areal unit problem (Openshaw 1984); and we did not rely on standard and spatial statistics with a well-defined mean to characterize spatial heterogeneity, but rather power-law-based statistics, driven by fractal and scaling thinking. Therefore, the underlying ways of thinking adopted in this study are (1) bottom up rather than top down, in terms of data and methods, and (2) nonlinear rather than linear, or fractal rather than Euclidean in terms of the power-law statistics.

The remainder of this paper is structured as follows. Section 2 presents the methods in which we define the concept of natural cities, and discuss ways of characterizing natural cities. Section 3 presents the data on a monthly basis and shows basic statistics of the data. Section 4 discusses on the results and major findings, while Section 5 on the implications of the study. Finally, Section 6 draws a conclusion and points to future work.

**2. Methods**
In this section, we illustrate and define the concept of natural cities and present various ways of characterizing natural cities. We also discuss how natural cities differ from conventional cities and why the notion of natural cities represents a new way of thinking for geospatial analysis.

*2.1 Defining natural cities*
To approach the difficult task of defining and describing natural cities, we start with definitions of conventional cities and try to clarify why the conventional definitions are little natural. A city is a relatively large and permanent human settlement. But how large a settlement must be to qualify as a city is unclear. For example, a city in Sweden may not qualify as a city in China. Also, many cities have a particular administrative, legal, and historical status according to their local laws. In the United States, for example, cities can refer to incorporated places, urban areas, or metropolitan areas with sufficient population of, say, at least 10,000. This population threshold can be very subjective and is dependent on the country. This subjectivity is also demonstrated in the physical boundaries of cities, which are legally and administratively determined. Remotely sensed imagery provides new means to delineate city boundaries, but how does one choose an appropriate pixel value as a cutoff for the delineation? Because of these subjectivities, conventional definitions of cities are little natural. How, then, can we define a city in natural ways?

We present three examples of natural cities before formally define the concept. In the first example, natural cities are derived from massive street nodes, including both junctions and street ends. Given all street nodes of an entire country, we can run an iterative clustering algorithm to determine whether a node is within the neighbor of another node. For example, set a radius of 700 meters and continuously draw a circle around each node to determine whether any other node is within its circle. This progressive and exhaustive process results in many natural cities; see Figure 1a for an illustrative example. In their study, Jiang and Jia (2011) found that millions of natural cities could be derived from dozens of millions of street nodes in the United States using OpenStreetMap (OSM) data (Bennett 2010). Instead of massive street nodes, the second example relies on a massive number of street blocks to extract natural cities. Jiang and Liu (2012) adopted the three largest European countries: France, Germany, and the UK for their case studies, again using OSM data. The idea is illustrated in Figure 1b in which small blocks (smaller than a mean) constitute a natural city. Although this method sounds very simple, the computation is very intensive for each country, involving millions of street blocks. The third example comes from Jiang and Yin (2014), in which the authors relied on nighttime imagery to derive natural cities. The authors took all pixel values (millions of pixels each valued between 0 and 63) of an image in the United States and computed an average or mean. The mean split all the pixels into two: those above the mean, and those below the mean. For the pixels above the mean, a second mean was obtained, and it can be a meaningful cutoff for delineating natural cities.

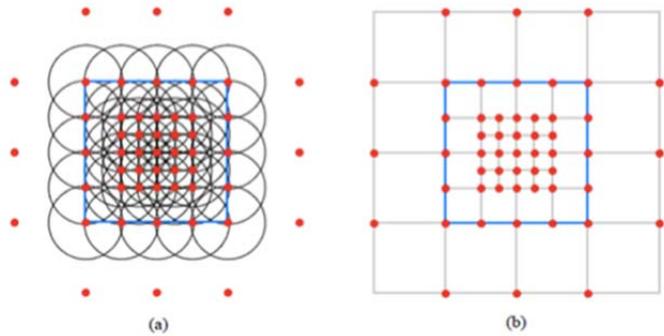

Figure 1: (Color online) Natural cities based on (a) street nodes and (b) street blocks
(Note: Blue rectangles are the boundaries of the natural cities, which are composed of high-density nodes or small street blocks based on the head/tail division rule (Jia and Jiang (2010))

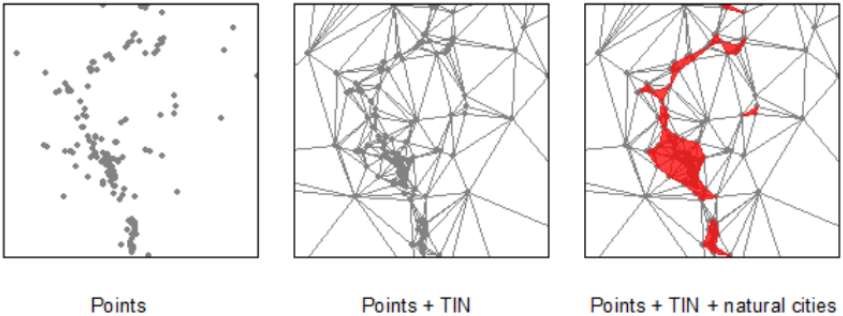

Figure 2: (Color online) Procedure of generating natural cities (red patches) from points through TIN

These examples of deriving natural cities point out the importance of the mean's effect, which is based on the head/tail division rule: *Given a variable X, if its values x follow a heavy tailed distribution, then the mean (m) of the values can divide all the values into two parts: a high percentage in the tail, and a low percentage in the head* (Jiang and Liu 2012). The heavy tailed distribution refers to the statistical distributions that are right-skewed, for example, power law, lognormal, and exponential. Obviously, the density of street nodes, the size of street blocks, and the nighttime imagery pixel values all exhibit a heavy tailed distribution, which implies that there are far more small things than large ones. In this paper, we build up a huge triangular irregular network (TIN) based on social media users' locations, and then categorize these small triangles (smaller than a mean) as natural cities (Figure 2); refer to the Appendix for a short tutorial. Section 5 further discusses why the head/tail division rule works so well in delineating natural cities.

Based on these examples, a formal definition of natural cities can be derived. Natural cities refer to human settlements or human activities in general on Earth's surface that are objectively or naturally defined and delineated from massive geographic information of various kinds, and based on the head/tail division rule. Unlike conventional cities, natural cities do not need to meet a minimum population requirement. A one-person settlement may constitute a natural city, or even zero people, if natural cities are defined not according to human population, but something else. For example, when natural cities are defined according to street nodes, a natural city derived from one street node may have no people there at all. The reader may question whether this definition makes sense. The definition makes good sense because it provides a new perspective for geospatial analysis, and helps us develop new insights into geographic forms and processes (see Sections 4 and 5). That is also the reason that we use the term natural cities to refer to human settlements or human activities in general on the Earth's surface. With the concept of natural cities, we abandon the top-down imposed geographic units or boundaries such as states, counties, and cities, in order to study geographic forms and processes more scientifically.

*2.2 Characterizing natural cities*
The rank-size distribution of cities in a region can be well characterized by Zipf's law, i.e., an inverse power relationship between city rank (r) and city size (N), $N = r^{-1}$ (Zipf 1949). Simply put, when ranking all cities in a decreasing order for a given country, the largest city is twice as big as the second largest, three times as big as the third largest, and so on. In other words, a city's size by population is inversely proportional to its rank. Such a simple and neat law is found to hold remarkably well for almost all countries or regions (e.g., Berry and Okulicz-Kozaryn 2011), although some researchers have challenged its universality (e.g., Benguigui and Blumenfeld-Leiberthal 2011). Essentially, Zipf's law indicates two aspects: (1) a power-law relationship between rank and size, and (2) the Zipf's exponent of one. Most previous studies have confirmed the first aspect, but not the second; the Zipf's exponent was found to deviate from one. In other words, the first aspect is not as much controversial as the second aspect. Some researchers argued that Zipf's law was primarily used for characterizing large cities rather than all cities. In this study, we chose large natural cities (larger than a mean) to examine whether they followed Zipf's law. The scaling patterns of far more small cities than large ones underlie Zipf's law — a majority of small cities, while a minority of large cities. More important, the scaling pattern recurs not just once, but multiple times for those large cities, again and again. This is the basis of head/tail breaks (Jiang 2013), a novel classification scheme for data with a heavy tailed distribution. In what follows, we illustrate head/tail breaks with a working example.

Table 1: Head/tail breaking statistics for the TIN edges

| Edges | Mean | # Head | % Head | # Tail | % Tail |
|---|---|---|---|---|---|
| 504 | 2.2 | 135 | 27% | 369 | 73% |
| 135 | 6.2 | 35 | 26% | 100 | 74% |
| 35 | 13.4 | 13 | 37% | 22 | 63% |
| 13 | 20.7 | 3 | 23% | 10 | 77% |
| 3 | 33.2 | 1 | 33% | 2 | 67% |

The triangulated irregular network shown in Figure 2 apparently seems to contain far more short edges than long ones. This is indeed true. There are 504 edges, ranging from the shortest 0.001 to the longest 46.752. The wide range 46.751 = 46.752 – 0.001 and the large ratio 46,752 = 46.752/0.001 clearly indicate far more short edges than long ones, although the exact distribution is up to further statistic investigation. The average length of the 504 edges is 2.2, which splits all the edges into two unbalanced parts: 135 in the head (27 percent) and 369 in the tail (73 percent). This head/tail breaking process can be continued for the head again and again, as shown in Table 1. Eventually, the scaling pattern of far more short edges than long ones recurs five times, three of which are plotted in Figure 3, or so-called nested rank-size plots. Given that the scaling pattern recurs five times, the ht-index is six. Note that ht-index (Jiang and Yin 2014) is an alternative index to fractal dimension (Mandelbrot 1983) used to capture the complexity of geographical features.

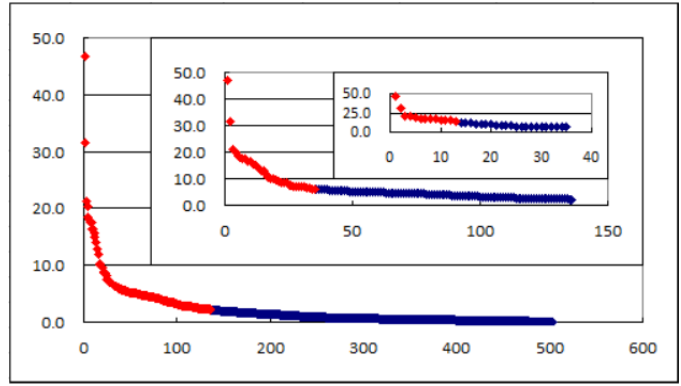

Figure 3: (Color online) Nested rank-size plots for the first three hierarchical levels with respect to the first three rows in Table 1
(Note: The x axis and y axis represent rank and size respectively. The largest plot contains the 504 edges, the red being the first head (135 edges) and the blue being the first tail (369 edges). The 135 edges are plotted again with the red representing 35 in the second head and the blue 100 in the second tail. The smallest plot is for the 35 edges in the second head.)

Head/tail breaks or ht-index provides a simple yet effective means to characterize natural cities, or data in general with a heavy tailed distribution for mapping purposes. The derived ht-index captures the hierarchy or scaling hierarchy of the data. For mapping purposes, head/tail breaks is superior to conventional classification methods for capturing the underlying scaling pattern (Jiang 2013). Ht-index complements to fractal dimension for characterizing the complexity of geographic features or fractals in general.

**3. Data and Data Processing**
As stated above, the data for this study came from the former location-based social medium Brightkite, during its three-year (31 months to be more precise) life span, from April 2008 to October 2010 (Cho, Myer, and Leskovec 2011). The case included 2,788,042 locations in the mainland United States. From the amount of locations, we removed duplicate locations, obtained 404,174 unique locations for generating a TIN, and then 8,106 natural cities as of October 2010, by following the procedure shown in Figure 2, as well as the short tutorial presented in the Appendix. The location data was time stamped (Table 2), so we were able to slice all these locations monthly in an accumulated manner, i.e., locations at month $m_{i+1}$ contain all locations between months $m_1$ and $m_i$, where $1 \leq i \leq 31$. For each time interval or snapshot, we generated a set of natural cities ranging from 100 to 8106. Figure 4 illustrates the 8,106 natural cities as of October 2010, showing their boundaries and populations. Note that this is just one of the 31 snapshots or datasets in the study.

Table 2: Initial check-in data format

| User | Check-in time | Latitude | Longitude | Location id |
|---|---|---|---|---|
| 58186 | 2008-12-03T21:09:14Z | 39.633321 | -105.317215 | ee8b88dea22411 |
| 58186 | 2008-11-30T22:30:12Z | 39.633321 | -105.317215 | ee8b88dea22411 |
| 58186 | 2008-11-28T17:55:04Z | -13.158333 | -72.531389 | e6e86be2a22411 |
| 58186 | 2008-11-26T17:08:25Z | 39.633321 | -105.317215 | ee8b88dea22411 |
| 58187 | 2008-08-14T21:23:55Z | 41.257924 | -95.938081 | 4c2af967eb5df8 |

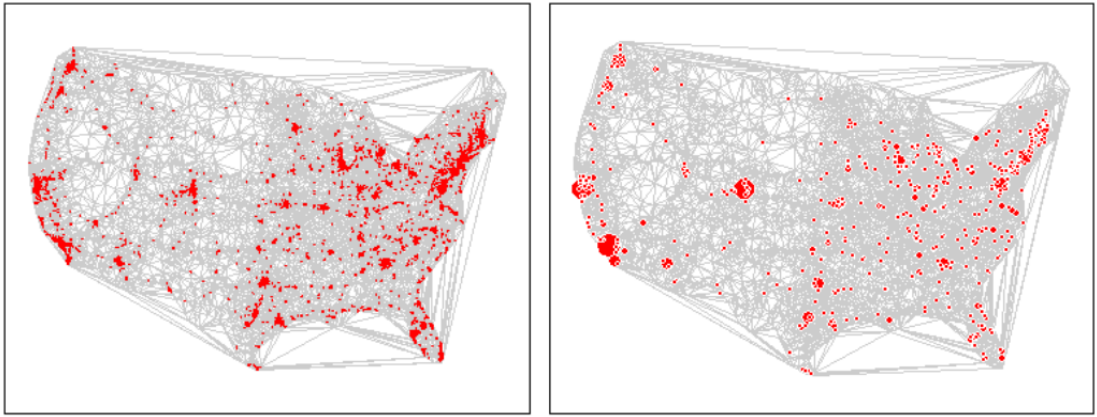

Figure 4: (Color online) The largest set of natural cities as of October 2010 (red patches for boundaries and red dots for populations) on the background of TIN (gray lines) generated from 404,174 unique location points or 2,788,042 duplicate ones

Table 3: Measurements and statistics from location points to natural cities for the different time intervals
(Note: Pnt = # of points, PntUniq = # of unique points, TINEdge = # of TIN edges, Mean = Average length of TIN edges, NaturalCity = # of natural cities)

| Time | Pnt | PntUniq | TINEdge | Mean | NaturalCity |
|---|---|---|---|---|---|
| **2008-04** | **8,563** | **3,007** | **9,001** | **33,775** | **100** |
| **2008-05** | **104,848** | **21,171** | **63,492** | **10,589** | **437** |
| 2008-06 | 207,048 | 35,176 | 105,505 | 7,910 | 698 |
| 2008-07 | 308,987 | 46,447 | 139,317 | 6,882 | 938 |
| **2008-08** | **405,866** | **57,458** | **172,351** | **6,129** | **1,147** |
| ... | | | | | |
| **2008-12** | **903,990** | **135,998** | **407,970** | **3,636** | **2,558** |
| ... | | | | | |
| **2009-06** | **1,800,825** | **263,103** | **789,284** | **2,518** | **5,132** |
| ... | | | | | |
| **2009-12** | **2,407,118** | **361,990** | **1,085,945** | **2,109** | **7,243** |
| ... | | | | | |
| **2010-10** | **2,788,042** | **404,174** | **1,212,498** | **1,969** | **8,106** |

Table 3 lists some basic measurements and statistics from the location points to the natural cities. For example, for the first month, April 2008, only 100 natural cities were generated from 8,876 locations, of which 3,007 unique locations were used for generating a TIN with 9,001 edges, and a mean of 33,775 as the cutoff to derive the 100 natural cities. The number of natural cities increased steadily to 8,106 as of October 2010. In the following section, we utilize the seven time intervals highlighted in Table 3 for a detailed discussion of our findings.

## 4. Results and Discussion

Before discussing the findings, we map the natural cities at the seven time intervals (or snapshots) for four largest natural cities surrounding Chicago, New York, San Francisco, and Los Angeles. These are shown in Figure 5, which illustrates clearly how the four regions grew during the 31-month period. All parts of the country can be assessed for similar patterns of growth and evolution. We know little why the procedure illustrated in Figure 2 and the Appendix, works so well, but the resulting patterns suggest that the natural cities effectively capture the evolution of real cities. On the one hand, the natural cities expanded towards more fragmented pieces, far more small pieces than large ones. On the other hand, the physical boundaries of the natural cities tended to become rougher over time. These two aspects suggest that the natural cities are fractal, and become more and more fractal, resembling very well real cities (Batty and Longley 1994). These two aspects are further discussed in the following.

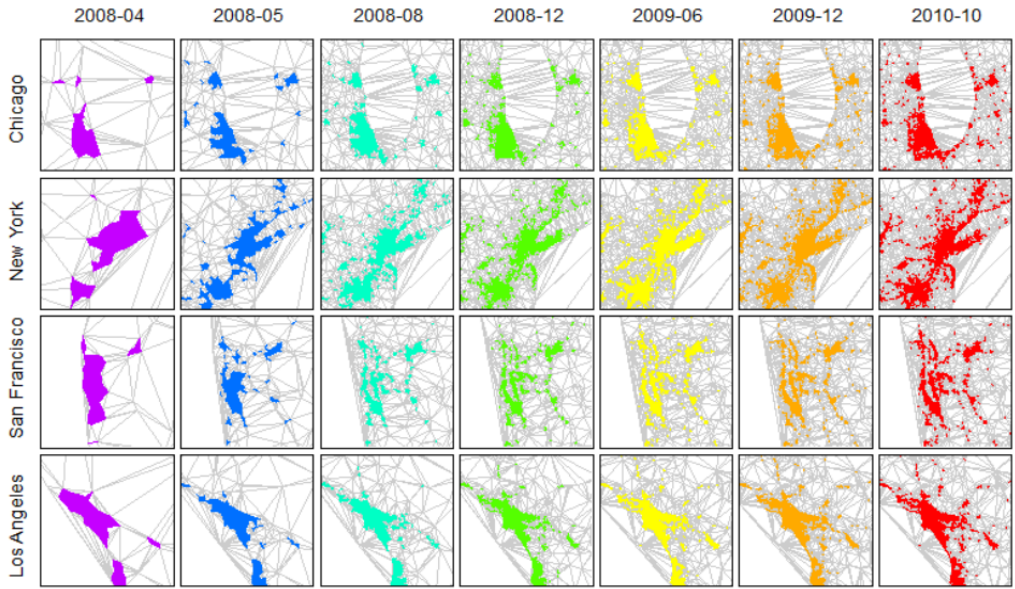

Figure 5: (Color online) Evolution of the natural cities near the four largest cities regions with TIN as a background

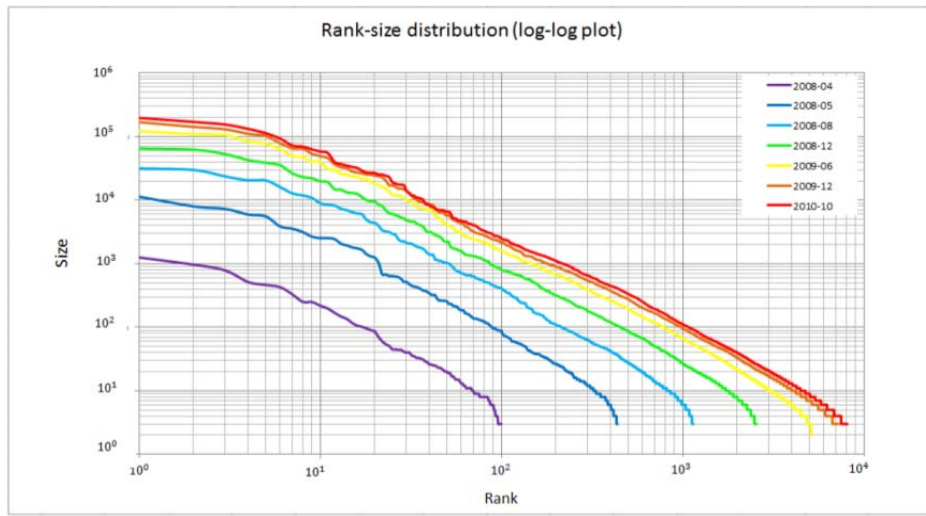

Figure 6: (Color online) Rank-size plot for the natural cities

These results can be assessed from both global and local perspectives. Globally, all the natural cities in the United States exhibit a power-law distribution. This is shown rank-size plots (Figure 6), in which the distribution lines are very straight for all the natural cities at the different time intervals in the log-log plots. The natural cites as of April 2008, except the smallest with less than 12 people, exhibited a clear power law, probably the straightest distribution among all others. However, the distribution lines from May 2008 to October 2010 are less straight, indicating that a few of the largest natural cities did not fit well the power-law distribution. This is particularly obvious for the last two snapshots in December 2009 and October 2010. A possible reason for this difference, moving from a striking to a less striking power law, is described below.

In further examinations, we looked at the large cities (larger than the mean) in each snapshot and found that Zipf's exponent was indeed around one for the first month (0.99), and then greater than one by about 0.3 (Table 4). Considering the duality of Zipf's law, this result suggests that Zipf's law held remarkably well for the first month, but less so for the remaining months. We postulated a possible reason: The social medium users at the first month increased proportionally with the populations of real cities, thus leading to a striking Zipf's law effect among the natural cities because the populations of real cities are power-law distributed. Over time, large cities — particularly a few of the largest cities such as New York — did not capture the other cities in attracting more users. In other words, beyond the first month, the increase in social medium users became less proportional to the real cities' populations. As a result, Zipf's law is less striking. In contrast to small deviations of Zipf's exponent, the ht-index increased from 2 to 5 (Table 4). The increment of the ht-index implies that more hierarchical levels were added, reflecting well the evolution of the natural cities and of the social medium.

Table 4: Zipf's exponent and ht-index for the natural cities

|  | 2008-04 | 2008-05 | 2008-08 | 2008-12 | 2009-06 | 2009-12 | 2010-10 |
|---|---|---|---|---|---|---|---|
| Zipf's exponent | 0.99 | 1.19 | 1.25 | 1.28 | 1.29 | 1.28 | 1.26 |
| Ht-index | 2 | 4 | 4 | 4 | 4 | 5 | 5 |

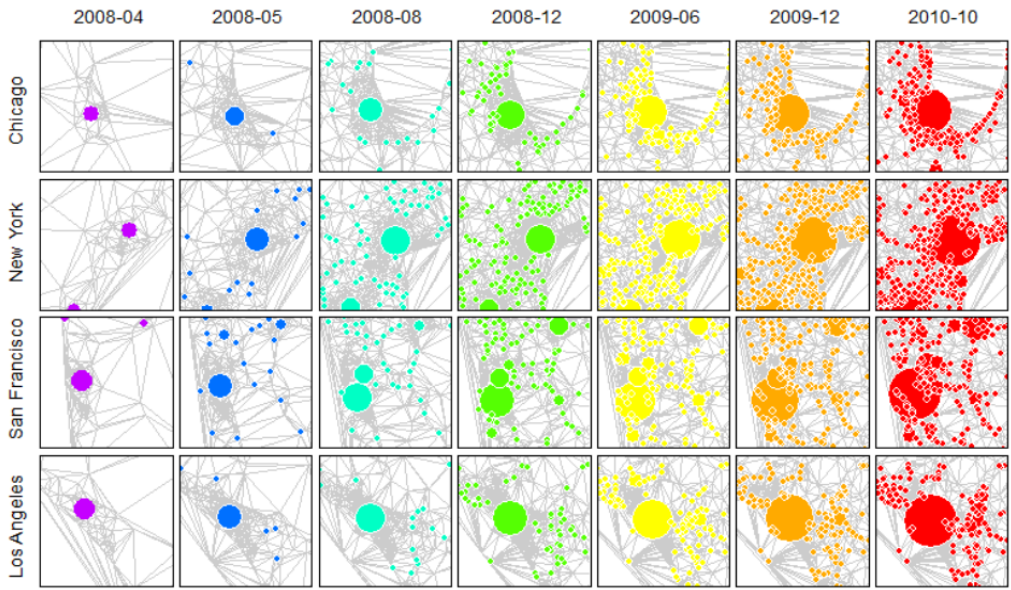

Figure 7: (Color online) Evolution of the natural cities in terms of populations (or points) near the four largest cities regions

Locally, there are two points to discuss. First, the boundaries of the natural cities became rougher over time, very much like the Koch curve when the iteration goes up. For example, the boundaries of the natural cities as of April 2008 were simple enough to be described by Euclidean geometry. However, over time, the boundaries must be characterized by fractal geometry — more fragmented with more fine scales added. Second, large natural cities tended to become larger and larger, while small ones continuously emerged at local levels. Figure 5 illustrates this finding in a less striking manner, as the city sizes are measured by the physical extents. But if the city sizes are measured by population as in Figure 7, we noticed the rapid increases for the four largest cities. Overall, the four cities tended to become larger and larger, but there was a major difference among the four. To illustrate the difference, we must clarify that Figure 7 adopts the graduated dots to represent the city sizes, which are classified according to head/tail breaks. This is because the city sizes exhibited a heavy tailed distribution, or there were far more small cities than large ones. Therefore, the dot sizes in Figure 7 do not represent city sizes, strictly speaking, but rather, the corresponding classes to which the cities belong. Notice that the largest natural city in the Chicago region in October 2010 appears smaller than others. This is indeed true! Table 5 clearly indicates that the Chicago natural city in October 2010 belongs to the sixth among the seven classes, while the others belong to the seventh among the seven classes.

Table 5: Evolution of the four cities within the system of the natural cities
(Note: a/b where a and b respectively denote the class the particular city belongs to, and the total number of classes or the ht-index)

|  | 2008-04 | 2008-05 | 2008-08 | 2008-12 | 2009-06 | 2009-12 | 2010-10 |
|---|---|---|---|---|---|---|---|
| Chicago | 3/4 | 4/5 | 4/5 | 5/6 | 5/6 | 5/6 | 6/7 |
| New York | 3/4 | 5/5 | 5/5 | 5/6 | 6/6 | 6/6 | 7/7 |
| San Francisco | 4/4 | 5/5 | 5/5 | 6/6 | 6/6 | 6/6 | 7/7 |
| Los Angeles | 4/4 | 5/5 | 5/5 | 6/6 | 6/6 | 6/6 | 7/7 |

The above results can be summarized by one word - nonlinearity, which is reflected in both spatial and temporal dimensions. Spatially, the natural cities were distributed heterogeneously or unevenly, i.e., there were far more small cities than large ones. This uneven distribution also was seen in the temporal dimension. For example, within the first 9 months of 2008, the natural cities had already been shaped (Figure 5), with populations continuously growing, and small natural cities being added persistently for the remaining time. In other words, it took just one third of the social medium's lifetime to determine the shapes of individual cities. That is also the reason that we chose the seven unequal time intervals to examine the evolution.

**5. Implications of the Study**
The location-based social media provide large amounts of location data of significant value for studying human activities in the virtual world, as well as on the Earth's surface. Nowadays, the social sciences — human geography in particular — benefit considerably from emerging social media data that are time-stamped and location-based. The ways of doing geography and social sciences are changing! The emerging big data harvested from social media, as well as from geospatial technologies, coupled with data-intensive computing (Hey, Tansley, and Tolle 2009) are transforming conventional social sciences into computational social sciences (Lazer et al. 2009). In this section, we discuss some deep implications of this study for geography and social sciences in general.

The notion of natural cities implies a sort of bottom-up thinking in terms of data collection and geographic units or boundaries. Conventional geographic data collected and maintained from the top down by authorities are usually sampled and aggregated, and therefore, are small-sized. On the other hand, new data harvested from social media are massive and individual, so they are called 'big data.' Time-stamped and location-based social media data, supported by Web 2.0 technologies and contributed by individuals through humans as sensors (Goodchild 2007), constitute a brilliant new data source for geographic research. Conventional geographic units or boundaries are often imposed from the top down by authorities or centralized committees, while natural cities are defined and delineated objectively in some natural manner, based on the head/tail division rule. This natural manner guarantees that we can see a true picture of urban structure and dynamics, and suggests the universality of Zipf's law. This true picture is fractal and can be illustrated in this example: Throw forcefully a wine glass on a cement ground, and it will very likely break into a large number of pieces. Like the natural cities, these glass pieces are fractal or follow Zipf's law: On the one hand, there are far more small pieces than large ones, and on the other hand, each piece has an irregular shape.

The evolution of natural cities demonstrates nonlinearity at both spatial and temporal dimensions, or equivalently from both static and dynamic points of view. Many phenomena in human geography, as well as in physical geography, bear this nonlinearity (Batty and Longley 1994, Frankhauser 1994, Chen 2009, and Phillips 2003). However, we are still very much constrained by linear thinking, explicitly or implicitly, consciously or unconsciously. For example, we rely on Euclidean geometry to describe Earth's surface, and on a well-defined mean to characterize spatial heterogeneity. Our mindsets apparently lag behind the advances of data and technologies. Conventional linear thinking is not suitable for describing the Earth's surface (the geographic forms), not to mention uncovering the underlying geographic processes. Instead, we should adopt nonlinear thinking, or nonlinear mathematics such as fractal geometry, chaos theories, and complexity for geographic research. The tools adopted in this study, such as head/tail division rule, head/tail breaks, and ht-index, underlie nonlinear mathematics and power-law-based statistics. These nonlinear mathematical tools help to elicit new insights into the evolution of natural cities. Nonlinearity also implies that geographic forms and processes are unpredictable like long-term weather or climate in general. To better predict and understand geographical phenomena, we must seek to uncover the underlying mechanisms through simulations rather than simple correlations.

The head/tail division rule is intellectually exciting because it appears to be both powerful and mysterious. The reason why the head/tail division rule is an effective tool to derive natural cities, in particular at the different time stages, remains an open question. However, we tend to believe it is the effect of the wisdom of crowds — the diverse and heterogeneous many are often smarter than the few, even a few experts (Surowiecki 2004). The massive amount of edges (up to 1,212,498) of the generated TIN from the massive location points constituted the 'crowds,' and they collectively decided an average cutoff for delineating the natural cities. Every single edge had 'its voice heard' in the democratic decision. From the effectively derived natural cities, we can see an advantage of working with big data. If we had not worked with the entire US data set, but only an area surrounding New York for example, we would not have been able to determine a sensible cutoff for delineating the New York natural city. Only with the big data that includes all location points or all edges can a meaningful cutoff be determined and applied to all. In this sense, the approach to delineating natural cities is holistic and bottom up, with participation of all diverse and heterogeneous individuals.

It is important to note that the check-in users are biased towards certain types of people. Thus the derived natural cities are not exactly the same as the corresponding real cities. However, no one can deny that the boundaries shown in Figure 5 are not those of Chicago, New York, San Francisco, and Los Angeles. Note that this paper is not to study real cities, at an individual level, on how they can be captured or predicted by the natural cities, but to understand, at a collective level, underlying mechanisms of agglomerations, formed either by people in physical space (real cities), or by the check-in users in virtual space (natural cities). In other words, we consider cities (either real or natural cities) as an emergence (Johnson 2002) developed from the interaction of individual people, or that of cities. The fractal structure and nonlinear dynamics illustrated appear to be applied to both real and natural cities. The fact that not all people are the check-in users should not be considered a biased sampling issue. Sampling is an inevitable technique at the time of information scarcity, so called the small data era, but it is not a legitimate concept in the big data era. The large social media data implies N=all (Mayer-Schonberger and Cukier 2013). This N=all is an essence of big data. Given the 2.8 millions of check-in locations, the social media can be a good proxy for studying the evolution of real cities in the country.

We face an unprecedented golden era for geography, or social sciences in general, with the wave of social media and, in particular, the increasing convergence of social media and geographic information science (Sui and Goodchild 2011). For the first time in history, human activities can be documented at very fine spatial and temporal scales. In this study, we sliced the data monthly, but we certainly could have done so weekly, daily, and even hourly. We believe that the observed nonlinearity at the temporal dimension would be even more striking. This, of course, warrants further study. Geographers should ride the wave of social media and develop a more computationally minded geography or computational geography (Openshaw 1998). If we do not seize this unique opportunity, we may risk being purged from the sciences. The rise of computational social science is a timely response to the rapid advances of data and technologies. In fact, physicists and computer scientists already have been working on this exciting and rapidly changing domain (see Brockmann, Hufnage, and Geisel 2006; and Zheng and Zhou 2011). We geographers should do more rather than less.

## 6. Conclusion

Driven by the lack of data for tracking the evolution of cites, this study demonstrated that emerging location-based social media such as Flickr, Twitter, and Foursquare can act a proxy for studying and understanding underlying evolving mechanisms of cities. Compared with conventional census data that are usually sampled, aggregated, and small, the time-stamped and location-based social media data can be characterized as all, individual, and big. In this paper, we abandoned conventional definitions of cities, and adopted objectively or naturally defined natural cities, using massive geographic information of various kinds, and based on the head/tail division rule. Built on the notion of the wisdom of crowds, the head/tail division rule works very well to establish a meaningful cutoff for delineating natural cities. Natural cities provide an effective means or unique perspective to study human activities for better understanding of geographic forms and processes.

We examined the evolution of natural cities, derived from massive location points of the social medium Brightkite, during its 31-month life span. We found nonlinearity during the evolution of natural cities in both spatial and temporal dimensions, and the universality of Zipf's law. We archived all the data that could be of further use for developing and verifying urban theories. This study has deep implications for geography and social sciences in light of the increasing amounts of data that can be harvested from location-based social media. Therefore, we call for the application of nonlinear mathematics, such as fractal geometry, chaos theories and complexity to geographic and social science research. A limitation of this study lies in the data that shows only the social medium's continuous rise and not its decline. Brightkite seemed to disappear overnight. Future research should concentrate on development of power-law-based statistics, and underlying nonlinear mathematics, to manage the increasing social media data and on agent-based simulations to reveal the mechanisms for the evolution of natural cities.

**Acknowledgement**
An early version of paper was presented by the first author as a keynote entitled "The evolution of natural cities: a new way of looking at human mobility", at Mobile Ghent '13, 23-25 October 2012, University of Ghent, Belgium. We would like to thank Kuan-Yu Huang for partial data processing, the three anonymous referees and editor for their comments that significantly improved the quality of this paper.

**Appendix: Tutorial on How to Derive Natural Cities based on ArcGIS**

This tutorial aims to show, in a step by step fashion with ArcGIS, how to derive natural cities using the first month data (2008-04) as an example. Once you have got the check-in data of the first month, transfer them into an Excel sheet with two columns namely x, y, respectively representing longitude and latitude. Add a third column z, and set all column values as one (or any arbitrary value since ArcGIS relies on 3D points for creating a TIN). Insert the Excel sheet as a shape file data layer in ArcGIS (Figure A1(a)). Create a TIN from the point layer, using ArcToolbox > 3D analyst tools > TIN management > Create TIN (Figure A1(b)).

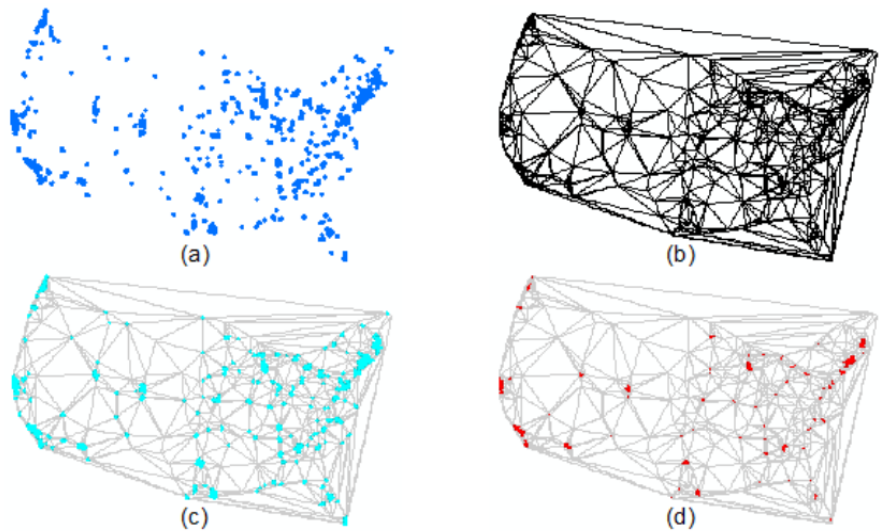

Figure A1: Screen snapshots including (a) the 3,007 unique points, (b) the TIN from the 3,007 points, (c) the selected edges shorter than the mean 33,775, and (d) the 100 natural cities created

Convert the TIN into TIN edge, using ArcToolbox > 3D analyst tools > Conversion > From TIN > TIN Edge. The converted TIN edge is a polyline layer. Right click the polyline layer to Open Attribute Table in order to get statistics about the length of the edges. Figure A2 shows that the frequency distribution that is apparently L-shaped, indicating that there are far more short edges than long ones. Note that the mean is 33,775.

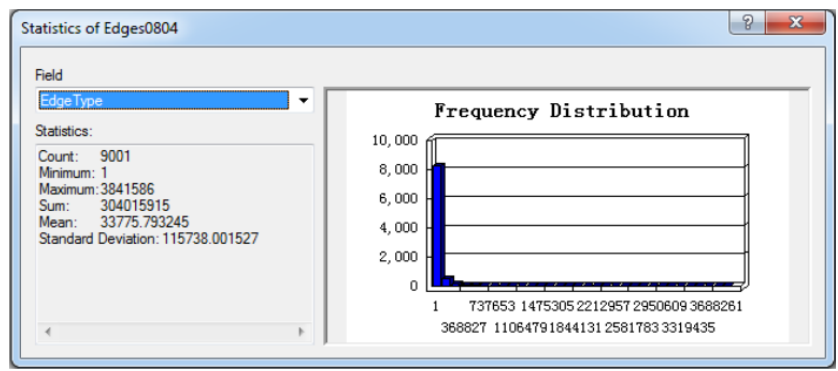

Figure A2: Statistics of TIN edges

Select those shorter edges than this mean 33,775 following menu Selection > Select by Attributes... . The selected edges are highlighted in Figure A1(c). The selected shorter edges refer to high density locations. Dissolve all the shorter edges into polygons to be individual natural cities (Figure A1(d)), following ArcToolbox > Data Management Tools > Features > Feature to Polygon > Generalization > Dissolve, or alternatively following menu Geoprocessing > Dissolve (the option 'Create multipart features' should be unchecked).

The above steps all can be done with the existing ArcGIS functions. The reader may encounter some problem with more check-in points are added cumulatively. In this case, we must spit the data into small pieces and put them back again into ArcGIS with some simple codes. All the produced data are archived at https://sites.google.com/site/naturalcitiesdata/.